\documentclass[conference]{IEEEtran}
\usepackage[T1]{fontenc}
\usepackage[utf8]{inputenc}
\usepackage{lmodern}
\usepackage{amsmath}
\usepackage{booktabs}
\usepackage{tikz}
\usetikzlibrary{arrows.meta,positioning,fit}
\usepackage{url}
\usepackage[hidelinks]{hyperref}

\begin{document}

\title{An Open Implementation Flow for the Xilinx Spartan-3E and Spartan-6 that Runs in the Browser,\\
Checked Byte for Byte Against ISE}

\author{\IEEEauthorblockN{Pedro Mal\'o}
\IEEEauthorblockA{Departamento de Engenharia Electrot\'ecnica e de Computadores (DEEC)\\
Faculdade de Ci\^encias e Tecnologia (FCT)\\
Universidade NOVA de Lisboa, Portugal}}

\maketitle

\begin{abstract}
The author describes an open implementation flow, from HDL to a programmable bitstream, for the Xilinx Spartan-3E
(XC3S100E to XC3S1200E) and Spartan-6 (XC6SLX16, XC6SLX45), devices that only the discontinued ISE~14.7
supports. The flow runs unchanged in a web browser or in Node.js: Yosys and nextpnr compiled to
WebAssembly, with nextpnr's Himbaechel framework given Spartan-3E and Spartan-6 architectures generated
from Silinx's own device descriptions, bit database and timing model, and Silinx's own bitstream writer. The
bit database was built by black-box observation of ISE's documented XDL and \texttt{bitgen} tools, using
codeword-batched experiments; the router is given only switches whose bits were measured, and the writer
refuses a feature it has not measured. ISE is the oracle: for the same placed and routed design, the
bitstream is byte-identical to the one \texttt{bitgen} writes and ISE's DRC is clean. Placement and routing
were first done by an in-house annealer and PathFinder router, which nextpnr replaced once it matched them.
On 67 open-source designs for the XC3S1200E, compared with ISE at the same optimisation goal (45 designs
built by both), the flow uses 15\% fewer slices and has 14\% shorter critical paths at goal Area, 21\% fewer
slices and 8\% shorter paths at Speed, and 20\% fewer slices but 12\% longer paths at Fast (geometric
means). XST cannot read 15 of the designs. The flow is published as the npm package \texttt{silinx-fpga}
(Apache-2.0).
\end{abstract}


\section{Introduction}

Open FPGA flows exist for the Lattice iCE40 (IceStorm)~\cite{icestorm}, the ECP5 (Trellis)~\cite{trellis}
and the Xilinx 7-series (X-Ray)~\cite{xray}, among others, built on Yosys~\cite{yosys} and
nextpnr~\cite{nextpnr} or VPR~\cite{vtr}. Older Xilinx families have been left out, although many of their
boards are still in use. The Spartan-3E is a case in point: it is the device of the Digilent Basys2 and
Nexys2 teaching boards, it is supported only by ISE~14.7, unchanged since 2013, and Vivado does not support
it; the same holds for the Spartan-6 of the Nexys3 and Atlys.

This paper describes the implementation flow of Silinx, an open (AGPL-3.0) design environment for teaching
digital systems that runs in the browser. Its contributions are:
\begin{itemize}
\item a bit database for four Spartan-3E and two Spartan-6 devices, obtained by black-box observation of
ISE's documented tools with codeword-batched experiments, and a writer that refuses any feature it has not measured
(Section~\ref{sec:bitdb});
\item compact device descriptions, small enough to ship to a browser (Section~\ref{sec:device});
\item nextpnr Himbaechel architectures for the Spartan-3E and Spartan-6, generated from that data and run
in WebAssembly (Section~\ref{sec:pnr}), after an in-house placer and router that they replaced
(Section~\ref{sec:replace});
\item a validation method with ISE as an oracle (Section~\ref{sec:valid}); and
\item a comparison with ISE at three optimisation goals on 67 open-source designs (Section~\ref{sec:qor}).
\end{itemize}

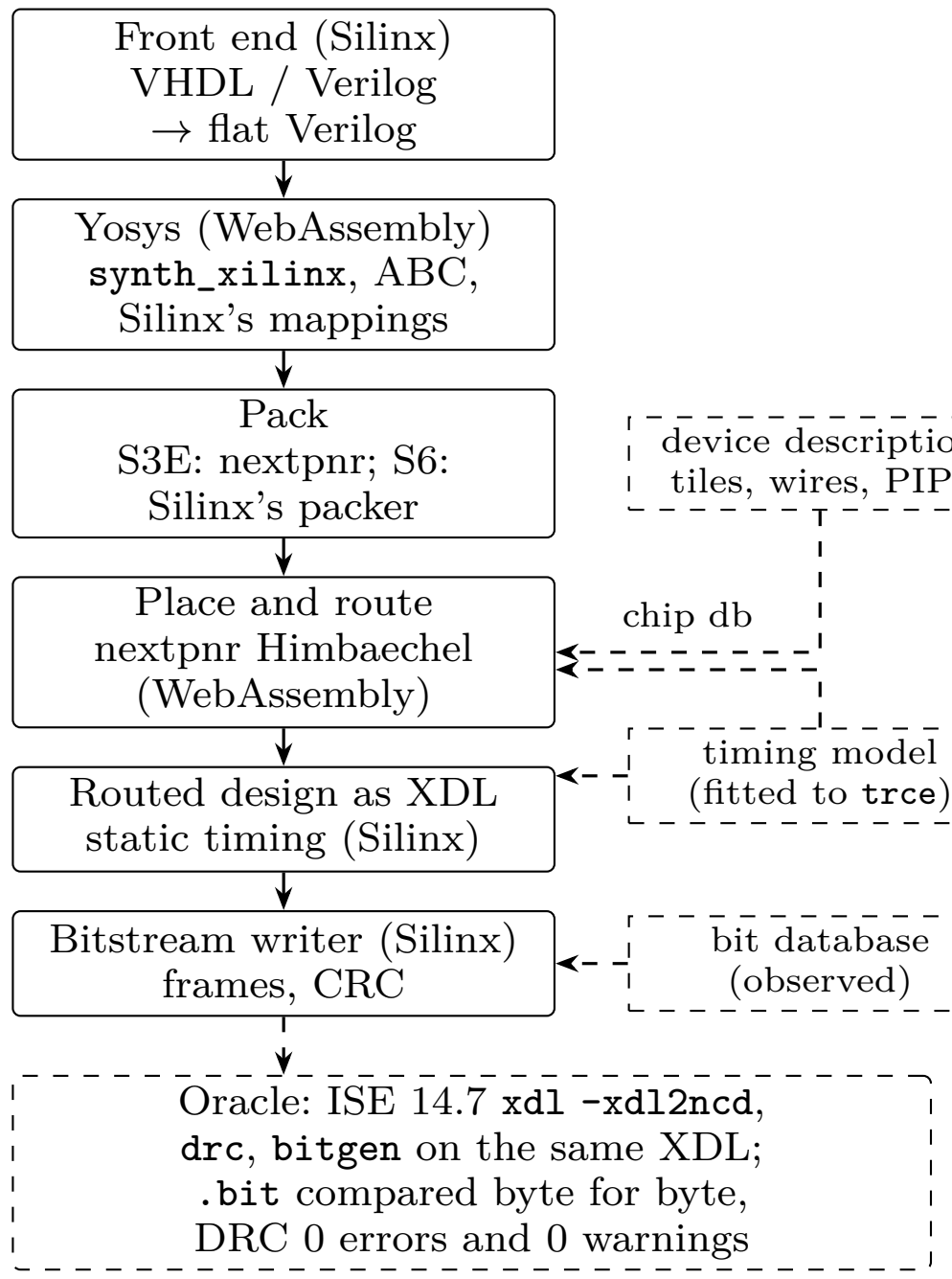
\begin{figure}[t]
\centering
\begin{tikzpicture}[node distance=2.6mm,
  box/.style={draw, rounded corners=1.5pt, align=center, font=\footnotesize, text width=36mm,
              minimum height=6mm, inner sep=2pt},
  data/.style={draw, dashed, align=center, font=\scriptsize, text width=25mm, inner sep=2pt},
  arr/.style={-{Stealth[length=1.6mm]}, semithick}]
\node[box] (fe) {Front end (Silinx)\\VHDL / Verilog $\to$ flat Verilog};
\node[box, below=of fe] (ys) {Yosys (WebAssembly)\\\texttt{synth\_xilinx}, ABC, Silinx's mappings};
\node[box, below=of ys] (pk) {Pack\\S3E: nextpnr; S6: Silinx's packer};
\node[box, below=of pk] (pr) {Place and route\\nextpnr Himbaechel (WebAssembly)};
\node[box, below=of pr] (xd) {Routed design as XDL\\static timing (Silinx)};
\node[box, below=of xd] (bg) {Bitstream writer (Silinx)\\frames, CRC};
\foreach \a/\b in {fe/ys,ys/pk,pk/pr,pr/xd,xd/bg} \draw[arr] (\a) -- (\b);
\node[data, right=5mm of pk] (dev) {device description\\tiles, wires, PIPs};
\node[data, right=5mm of xd, yshift=3mm] (sta) {timing model\\(fitted to \texttt{trce})};
\node[data, right=5mm of bg] (db) {bit database\\(observed)};
\draw[arr, dashed] (dev.south) |- node[pos=0.75, above, font=\scriptsize] {chip db} (pr.east);
\draw[arr, dashed] (sta.north) |- ([yshift=-1.2mm]pr.east);
\draw[arr, dashed] (sta.west) -- (sta.west -| xd.east);
\draw[arr, dashed] (db.west) -- (bg.east);
\node[box, dashed, below=4mm of bg, text width=62mm, xshift=13mm] (ise)
  {Oracle: ISE~14.7 \texttt{xdl -xdl2ncd}, \texttt{drc}, \texttt{bitgen} on the same XDL;\\
   \texttt{.bit} compared byte for byte, DRC 0 errors and 0 warnings};
\draw[arr, dashed] (bg.south) -- (bg.south |- ise.north);
\end{tikzpicture}
\caption{The flow. All stages run in Web Workers in the browser or in Node.js. The chip databases of
nextpnr are generated from Silinx's device description, keeping only the switches whose bits are measured,
with the delays of Silinx's timing model.}
\label{fig:flow}
\end{figure}

\section{Bitstream and Bit Database}
\label{sec:bitdb}

\subsection{Bitstream structure}
The Spartan-3E packet structure follows Xilinx's UG332: a sync word, register writes (frame length,
options, IDCODE), one Type~2 FDRI write of all frames, and a CRC. A frame holds $160 + 64r$ bits for $r$
CLB rows (frame counts in Table~\ref{tab:dev}). The CRC is a CRC-16 (polynomial 0x8005, bit-reversed) over
the 32 data bits followed by the 5 register-address bits of every register write, reset by RCRC and by
each check word; with the CRC disabled both check words are the constant \texttt{DEFC}.

\subsection{Method}
As in the projects above, the database records which bits each \emph{feature} sets: a LUT memory bit, a
slice setting (e.g.\ \texttt{CYSELF:F}), a pad's direction and standard, a routing switch (PIP). Test
designs are written in XDL, ISE's documented text format, converted by \texttt{xdl -xdl2ncd}, written by
\texttt{bitgen}, and compared. No ISE executable or internal data file is examined; the legal basis
(Art.~5(3) of Directive 2009/24/EC) is documented in the repository.

Features combine by bitwise OR, with a few documented exceptions that clear bits; this makes batching
possible. Each PIP of a routed design is assigned a binary codeword of length $L$ and weight $w$; variant
$i$ of the design keeps the PIPs whose codeword has bit $i$ set. Each bit of the bitstream then follows the
OR of the codewords of the PIPs that set it, and a PIP's bits are those whose pattern across the $L$
variants matches its codeword. Twenty variants measured all 4\,072 PIPs of a course design.

\subsection{Stages and coverage}
The Spartan-3E database was built in stages, each validated before the next: one design that numbers all
2\,448 slices of the XC3S250E in their LUTs; slice settings on a harness routed once by ISE; routing
switches and pads (about 185 ISE runs); block RAM contents, multipliers, distributed RAM, DCMs and the clock
tree (about 1\,780 runs); the XC3S100E, XC3S500E and XC3S1200E, whose tiles share the XC3S250E's tile-type
bits (about 2\,250 runs); and the remaining DCM settings, side global buffers and IOB registers (about
2\,000 runs; 112 held-out designs byte-identical). Under x86 emulation on an Apple Silicon machine an ISE
run takes about one minute. After the routing stage, 59\,177 of the 64\,329 PIPs of one tile of each of the
XC3S250E's 85 tile types had measured bits (92\%).

Unmeasured PIPs are left out of the routing graph given to the router, and the writer returns the list of
unmeasured features it would need instead of a bitstream. Settings first measured as unreliable were
refused until measured properly: random IOB input delays were identical in only about 60\% of held-out
designs until each pad's delay was measured on its own (970 designs).

\section{Device Descriptions}
\label{sec:device}

\begin{table}[t]
\centering
\caption{Shipped Spartan-3E device descriptions (gzipped JSON)}
\label{tab:dev}
\footnotesize
\setlength{\tabcolsep}{1.8pt}
\begin{tabular}{@{}lrrrrrrr@{}}
\toprule
Part & Slices & Frames & Tiles & Wires & PIPs & Tmpl. & MB \\
\midrule
XC3S100E  &   960 & 369 &   667 &   250\,112 &   891\,252 & 266 & 1.64 \\
XC3S250E  & 2\,448 & 578 & 1\,419 &   569\,094 & 2\,139\,368 & 293 & 1.96 \\
XC3S500E  & 4\,656 & 730 & 2\,255 &   975\,322 & 3\,899\,604 & 294 & 2.19 \\
XC3S1200E & 8\,672 & 959 & 3\,710 & 1\,691\,954 & 6\,997\,166 & 321 & 2.54 \\
\bottomrule
\end{tabular}
\end{table}

Placement and routing need the tiles, sites, wires and PIPs. They are computed from ISE's documented
\texttt{xdl -report -pips -all\_conns} output (204\,MB of text for the XC3S250E, 595\,MB for the XC3S1200E).
Most tiles of a type are identical, so each distinct tile content is stored once as a \emph{template}
(wires, PIPs and site pins by local index), each tile names its template, and wires joined across tiles are
stored as varint-coded \emph{nodes}. For the XC3S250E, 576 CLB tiles share one template, and the file loads
in about 90\,ms. Descriptions are shipped only for the devices of the supported boards
(Table~\ref{tab:dev}, and the XC6SLX16 and XC6SLX45). The four Spartan-3E dies have also been described
purely by observation, using ISE as an oracle of existence and its DRC to decide whether wires join; the
examples routed on each measured description pass ISE's DRC with byte-identical bitstreams. The shipped
files are still the computed ones.

\section{Synthesis and Optimisation Goals}
\label{sec:synth}

Silinx's own VHDL/Verilog front end writes the elaborated design as one flat Verilog module for Yosys
(\texttt{synth\_xilinx -family xc3se}). The stock mapping gave 886 LUTs and 522 slices on a course design,
against XST's 309 and 180. Mapping with ABC~\cite{abc} onto LUT4s at their cost, narrow arithmetic as
tables, SRL16 inference and a second, depth-oriented mapping (SOP balancing) brought it to 278 LUTs and 165
slices. Further candidates were added later: ABC9 onto LUT4s, adder chains written as
gates rewritten onto MUXCY/XORCY, trees of 2:1 multiplexers mapped onto MUXF5--F7, state machines tried
one-hot, and wide decodes on the critical path collapsed to BDDs and mapped again.

Optimisation is chosen, as in ISE, by a goal:
\begin{itemize}
\item \emph{Fast}: the usual mapping alone and one placement. Since release 15.23.0 a design deeper than
16 levels of logic (at most 1\,500 LUTs) also gets Speed's synthesis, kept when it is 10\% faster by the
estimate.
\item \emph{Area} (the default of a new project): the smaller (by slices, then LUTs) of the usual mapping's
netlist and Speed's, and one placement.
\item \emph{Speed}: every mapping for speed, and several placements, each routed, the fastest kept by
Silinx's timing analysis (32 for designs of up to 100 slices, 8 above; since 15.23.0, 4 above 500 slices).
\end{itemize}

\section{Placement and Routing with nextpnr}
\label{sec:pnr}

nextpnr's Himbaechel framework takes an architecture as a generated chip database plus a C++
\emph{micro-architecture} for packing and placement rules. Silinx's architectures for the Spartan-3E and
Spartan-6 live in a fork of nextpnr~\cite{nextpnrspartan}; each is compiled to WebAssembly and published as
an npm package (at versions 1.4.1 and 0.1.3).

\subsection{Chip database}
A generator reads Silinx's device description: one tile type per template, the wires, PIPs and nodes, and
the BELs inside the sites, written by hand because the description has no site internals. PIPs whose bits
the database does not know are left out (28\,030 of 2\,108\,744 on the XC3S250E in the first build), so that
nextpnr cannot route on them. nextpnr counts a PIP's delay rather than a wire's, so each PIP carries Silinx's
delay for the class of the node it drives (double, hex, long, OMUX, IMUX, global clock, and so on). A LUT's
inputs are wires of their own, each reached from every site pin by a PIP carrying that pin's delay, so the
router puts critical signals on fast pins; the conversion back rewrites each LUT's equation for the pins
chosen.

\subsection{Micro-architecture}
The Spartan-3E micro-architecture takes every cell Silinx's own packer took: LUT--flip-flop pairs sharing a
slice's control set, carry chains, wide multiplexers in the CLB's fixed F5--F8 pattern, SLICEM memories and
shift registers, multiplier cascades, DCMs, IOB registers and DDR, latches, falling-edge flip-flops, clock
switches, and the global clock lines (a clock pad's buffer placed where its dedicated path reaches).
Constants that need no LUT are routed after placement from ISE-style dummy sources or a VCC site, and
inverters on a flip-flop's enable or set/reset become the slice's CEINV/SRINV. On the Spartan-6, Silinx's
own packer forms the slices first, and its rules place the pads, global buffers, block RAMs and DSPs, which
nextpnr is given as fixed.

Two changes to nextpnr's placer mattered for speed: the refinement after HeAP~\cite{heap} runs as an anneal
from a warm temperature (it had been a greedy pass), and the placer's delay prediction follows the driver
(a LUT, through an OMUX from a flip-flop) and the net's fanout, through a new architecture hook. Router~1's
effort is bounded, after which router~2 continues with alternative weights.

\subsection{Back to Silinx's tools}
nextpnr's routed design is converted to XDL: one instance per used site with its settings, each net's pins
and PIPs. Silinx's static timing and bitstream writer read that XDL, so the oracle of
Section~\ref{sec:valid} is unchanged. A research check simulates the converted XDL against nextpnr's netlist
cycle by cycle, comparing every flip-flop, because ISE's DRC cannot detect a wrong LUT equation.

\subsection{Timing model}
Wire delays were fitted by non-negative least squares to 10\,486 connections timed by \texttt{trce} (rms
0.099\,ns); cell arcs come from \texttt{trce} reports, including block RAMs and multipliers. Over 21 clocks
the estimate was within 1.4\% of \texttt{trce} on average (worst 4.0\%). The report also gives pad-to-setup,
clock-to-pad and pad-to-pad times, checks \texttt{OFFSET} constraints and derives DCM output periods.

\subsection{In the browser}
The chip databases are not shipped with the release: the first Place and Route of a device fetches them
(about 5\,MB for a Spartan-6 device) and the browser keeps them in its cache. The same back-end interface
also carries a third family without new reverse engineering: Lattice iCE40 designs build with the existing
tools, YoWASP's nextpnr-ice40 and \texttt{icepack}, downloaded on first use.

\section{Build, then Replace}
\label{sec:replace}

From 10 October 2026 the flow used its own placer and router. Placement was simulated annealing over slices
and rigid groups (carry chains, F6--F8 patterns) with VPR's schedule and a cost mixing half-perimeter
wirelength and criticality-weighted connection delays; routing was PathFinder~\cite{pathfinder}, with A*
from the whole net tree, two passes of critical rerouting and register replication. It placed a course
design of 777 LUTs in about 1.5\,s and routed it in under a second. Table~\ref{tab:history} gives its
comparisons with ISE (ISE at XST's default; Silinx at two synthesis efforts, Normal and High).

\begin{table}[t]
\centering
\caption{The in-house engine against ISE (geometric means of Silinx/ISE, 40--45 designs; below 0:
Silinx smaller or faster)}
\label{tab:history}
\footnotesize
\setlength{\tabcolsep}{3pt}
\begin{tabular}{@{}lrrrr@{}}
\toprule
 & \multicolumn{2}{c}{Normal} & \multicolumn{2}{c}{High} \\
\cmidrule(lr){2-3}\cmidrule(l){4-5}
Date (release) & LUTs & Crit.\ path & LUTs & Crit.\ path \\
\midrule
18 Oct (15.10.22) & $-16\%$ & $+20\%$ & $-20\%$ & $+5\%$ \\
20 Oct (15.11.2)  & $-18\%$ & $+3\%$  & $-20\%$ & $-2\%$ \\
25 Oct (15.13.0)  & $-20\%$ & $-7\%$  & $-22\%$ & $-9\%$ \\
\bottomrule
\end{tabular}
\end{table}

On 29 October a first Spartan-3E architecture for nextpnr placed and routed a counter (37 LUTs, 24
flip-flops) with a clean DRC and a byte-identical bitstream; its period was 8.43\,ns by \texttt{trce}, 7.15\,ns
once LUT pins could be permuted, against 6.72\,ns with Silinx's engine on the same netlist. The changes of
Section~\ref{sec:pnr} closed most of the gap: with the new placer refinement the best of four placements was
5.4\% slower than Silinx's engine on the examples (7.4\% before), and a later version shortened periods by
a further 1.4\% on average; the remaining gap was accepted. Routing constants and folding inverters as
Silinx's packer did brought the slice counts to Silinx's (traffic 66 against 66, vga 143 against 144).
nextpnr became the default with the old engine as a fallback in release 15.21.0, and the only engine in
15.22.0, on both families; register replication was removed with the old engine. On the Spartan-6, nextpnr's
speed gap to the old router was also accepted. The quality measured with nextpnr is reported next.

\section{Validation}
\label{sec:valid}

\paragraph*{Oracle}
\texttt{check-open-flow} writes the placed and routed design as XDL and has ISE convert it, run its DRC
(\texttt{drc -z}) and write a bitstream. Silinx's bitstream must be byte-identical to ISE's, with the CRC on
and off, and the DRC must report 0~errors and 0~warnings. This holds for the course examples on all six
devices, for feature fixtures (block RAM, multipliers, SRL16, DCMs, IOB registers, wide multiplexers,
constants, start-up options) and for the corpus designs checked; it runs in the nightly ISE tests. The
oracle proves the bitstream and the legality of the routing; it says nothing about the quality of the
placement, which Section~\ref{sec:qor} measures.

\paragraph*{Negative controls}
A design with all PIPs removed gives 372 DRC errors; a single dangling PIP gives an ``antenna'' error and no
bitstream; without the per-tile PIP exceptions, ISE refuses 51 PIPs in 11 of 12 designs.

\paragraph*{Hardware}
Bitstreams synthesised, placed, routed and written with no Xilinx tool run on a Basys2 (XC3S250E), among
them a nextpnr-routed counter whose blink rate matches the one built by Silinx's earlier engine.

\section{Quality of Results}
\label{sec:qor}

\paragraph*{Setup}
The corpus holds 67 open-source designs: CPUs (T80, light8080, PicoRV32, NEORV32, SERV, FemtoRV32),
peripherals (UART, SPI, I2C, FIFOs, IWLS'05 cores), cryptographic cores (AES, SHA-1, SHA-256, SipHash),
floating-point units and benchmarks (EPFL~\cite{epfl}, ITC'99, VTR~\cite{vtr}). Each is built by both flows
for the XC3S1200E-FG320 with the same top; a design with more ports than pins goes through the same I/O
wrapper in both. Silinx 15.22.1 (nextpnr architecture 1.4.0) builds each design at each goal; ISE~14.7 at the
matching XST setting: Area against \texttt{-opt\_mode Area -opt\_level 2}, Speed against \texttt{Speed}, level
2, and Fast against \texttt{Speed}, level 1, XST's default. ISE's map and par get their best chance at every
goal: \texttt{map -c 1} (the tightest packing) and a 5\,ns clock constraint, so that par works for speed.
Both critical paths are measured by \texttt{trce} on the routed designs (8 November 2026). Run times are each
flow's whole build, the Silinx flow in WebAssembly, and are only indicative.

\begin{table}[t]
\centering
\caption{Silinx 15.22.1 against ISE~14.7 at the same goal, XC3S1200E (geometric means of Silinx/ISE over the
45 designs built by both; below 0: Silinx smaller, faster or quicker)}
\label{tab:summary}
\footnotesize
\setlength{\tabcolsep}{3.2pt}
\begin{tabular}{@{}lrrr@{}}
\toprule
 & Area & Speed & Fast \\
ISE's XST setting & Area, 2 & Speed, 2 & Speed, 1 \\
\midrule
Logic LUTs              & $-16\%$ & $-21\%$ & $-19\%$ \\
Flip-flops              & $+5\%$  & $-9\%$  & $-4\%$  \\
Slices                  & $-15\%$ & $-21\%$ & $-20\%$ \\
As few LUTs or fewer    & 30/45 & 33/45 & 26/45 \\
\midrule
Critical path           & $-14\%$ & $-8\%$  & $+12\%$ \\
Designs timed by both   & 39 & 40 & 39 \\
As fast or faster       & 29/39 & 28/40 & 10/39 \\
Slower than ISE           & 10 & 12 & 29 \\
\quad their extra logic (ns)   & $+1.76$ & $-0.10$ & $+2.30$ \\
\quad their extra routing (ns) & $+1.15$ & $+1.00$ & $+1.28$ \\
\midrule
Run time (indication)   & $-80\%$ & $-11\%$ & $-84\%$ \\
\bottomrule
\end{tabular}
\end{table}

\begin{table*}[t]
\centering
\caption{Selected designs (XC3S1200E, 8 November 2026): slices Silinx/ISE, and critical paths by
\texttt{trce} in ns, at each goal against ISE at the matching XST setting. S/I below 1: Silinx smaller or
faster.}
\label{tab:qor}
\footnotesize
\setlength{\tabcolsep}{3.6pt}
\begin{tabular}{@{}lrrrrcrrrrcrrrr@{}}
\toprule
 & \multicolumn{4}{c}{Area} & & \multicolumn{4}{c}{Speed} & & \multicolumn{4}{c}{Fast} \\
\cmidrule(lr){2-5}\cmidrule(lr){7-10}\cmidrule(l){12-15}
Design & Slices S/I & Silinx & ISE & S/I & & Slices S/I & Silinx & ISE & S/I & & Slices S/I & Silinx & ISE & S/I \\
\midrule
epfl-adder        & 0.70 & 15.03 & 113.30 & 0.13 & & 0.44 & 10.07 & 11.70 & 0.86 & & 0.43 & 15.03 & 12.25 & 1.23 \\
epfl-priority     & 0.37 & 43.11 &  23.68 & 1.82 & & 0.46 &  8.86 & 21.38 & 0.41 & & 0.34 & 43.11 & 20.83 & 2.07 \\
epfl-max          & 0.80 & 68.16 & 117.47 & 0.58 & & 0.72 & 23.40 & 26.85 & 0.87 & & 0.59 & 68.16 & 25.68 & 2.65 \\
vtr-mkPktMerge    & 0.12 &  7.89 &  12.24 & 0.64 & & 0.12 &  6.76 &  8.47 & 0.80 & & 0.12 &  7.89 &  9.27 & 0.85 \\
light8080         & 0.41 & 11.94 &  15.38 & 0.78 & & 0.39 & 11.89 & 14.40 & 0.83 & & 0.40 & 11.78 & 15.13 & 0.78 \\
vtr-sha           & 0.76 & 15.08 &  19.50 & 0.77 & & 0.73 & 14.88 & 19.55 & 0.76 & & 0.69 & 15.08 & 19.92 & 0.76 \\
fpu-multiplier    & 0.58 & 14.04 &  14.05 & 1.00 & & 0.58 & 13.96 & 14.01 & 1.00 & & 0.61 & 13.92 & 14.61 & 0.95 \\
t80               & 1.06 & 17.71 &  20.49 & 0.86 & & 1.01 & 17.71 & 17.90 & 0.99 & & 1.15 & 23.48 & 18.13 & 1.29 \\
secworks-siphash  & 0.89 & 15.32 &  15.56 & 0.98 & & 0.94 & 15.13 & 13.53 & 1.12 & & 0.94 & 15.32 & 13.98 & 1.10 \\
iwls05-spi        & 0.86 & 15.37 &  13.22 & 1.16 & & 0.86 & 15.37 & 12.60 & 1.22 & & 0.96 & 15.05 & 13.04 & 1.15 \\
itc99-b05         & 2.10 & 12.81 &  11.63 & 1.10 & & 2.05 & 10.40 &  9.82 & 1.06 & & 2.00 & 12.81 &  9.82 & 1.30 \\
\midrule
Geometric mean    & 0.85 & \multicolumn{2}{r}{(39 designs)} & 0.86 & & 0.79 & \multicolumn{2}{r}{(40 designs)} & 0.92 & & 0.80 & \multicolumn{2}{r}{(39 designs)} & 1.12 \\
\bottomrule
\end{tabular}
\end{table*}

\paragraph*{Coverage}
Of the 67 designs, 45 are built by both flows at every goal. XST cannot read 15 of them (\texttt{\$clog2},
VHDL-2008 constructs such as reading an output port, libraries other than \texttt{work}, port type
mismatches, some loop and sensitivity-list forms), among them PicoRV32, NEORV32, SERV and the AES and SHA
cores; 7 more fail in ISE's constraint step on the I/O wrapper's pins. Silinx reads and implements all of
them, except epfl-multiplier at Area and Fast, where nextpnr 1.4.0 stopped on an error; version 1.4.1,
released the same day, builds it (3\,418 slices, 141.8\,ns), after the measurement. Designs without a
clocked path are not timed, which leaves 39 or 40 timed by both.

\paragraph*{Size}
At every goal Silinx uses 15--21\% fewer slices than ISE (Table~\ref{tab:summary}), with as few LUTs or fewer
on 26 to 33 of the 45 designs. The advantage comes largely from ABC's mapping, SRL16 inference and the
inverters folded into the slices. vtr-mkPktMerge is extreme (0.12 of ISE's slices) because ISE implements
its FIFOs as 153 SRL16s plus logic where Yosys infers far less. The opposite extreme is itc99-b05, a small
control circuit, at twice ISE's slices at every goal.

\paragraph*{Speed at Area and Speed}
At Area, Silinx's critical paths are 14\% shorter than ISE's at its Area setting, and as short or shorter
on 29 of 39 designs. This is partly because XST's area mode leaves deep logic deep: the EPFL adder (a
128-bit adder written as gates) takes 113.3\,ns with ISE, whose critical path crosses 127 LUTs, and 15.0\,ns
with Silinx, which maps it onto the carry chain. Silinx's Area keeps the smaller netlist even when it is slower: epfl-priority is 1.82 times ISE's
path at Area and 0.41 at Speed. At Speed, against ISE's \emph{Timing Performance} setting, the paths are 8\%
shorter (28 of 40 as short or shorter).

\paragraph*{Speed at Fast}
At Fast, Silinx's critical paths are 12\% longer than ISE's, and only 10 of 39 designs are as fast or
faster. Fast maps with the usual area-oriented mapping and makes one placement, and deep logic stays deep:
epfl-max takes 68.2\,ns over 57 levels of logic against ISE's 25.7\,ns over 17, epfl-priority 43.1 against
20.8\,ns. Fast is not the old engine's Normal of Table~\ref{tab:history} ($-7\%$): Normal kept a depth-oriented
mapping for deep designs, made several placements and replicated registers. Release 15.23.0 gave Fast the
depth rule back (Section~\ref{sec:synth}); it is not included in this measurement.

\paragraph*{Where the remaining gap is}
The comparison splits each critical path, as \texttt{trce} reports it, into logic (LUTs, carry,
clock-to-out and setup) and routing. On the designs slower than ISE, at Area (10 designs) and Fast (29) both
parts are longer, logic more than routing; at Speed (12 designs) the logic is as ISE's ($-0.10$\,ns) and the
difference is in the routing ($+1.00$\,ns). At this goal the gap therefore comes from nextpnr's placement
and routing on this architecture.

\paragraph*{Run time}
At Area and Fast a whole build takes 80--84\% less time than ISE's, in geometric mean; at Speed, with up to
32 placements, 11\% less. The largest designs remain slow in WebAssembly (secworks-aes and itc99-b17, about 9
and 11 minutes at Area).

\paragraph*{Against the best-known results}
Mapped onto 6-input LUTs, the EPFL combinational benchmarks give 1.82 times the size records and 1.95 times
the depth records of the EPFL competition (geometric means over the 17 of 20 the flow maps). Those records
came from SAT-based resynthesis and long optimisation runs; the flow synthesises in seconds.

\section{Spartan-6}
\label{sec:s6}
The same method covers the XC6SLX16 and XC6SLX45 of the Nexys3 and Atlys boards. The bitstream reader and
writer follow UG380 and \texttt{bitgen}'s packet sequence (16-bit configuration words, frames of 65 words, a
22-bit CRC). The LUT, slice (SLICEX, SLICEL, SLICEM), routing, clock management (DCM, PLL), block RAM, DSP48A1
and pad bits were measured with ISE, and the static timing model is fitted to \texttt{trce}, each
constraint within 5\% of its slack or 2.5\% of its period. Silinx packs the slices and nextpnr's Spartan-6
architecture places and routes them, in the local server and in the browser. The examples on both boards,
including a small processor running from block RAM, give byte for byte the bitstream \texttt{bitgen} writes
for the same routing, with no DRC error. DCM and PLL configurations beyond those measured are not covered
yet, and the quality of the Spartan-6 flow has not been compared with ISE on the corpus.

\section{Discussion}
\paragraph*{Effort of black-box documentation}
The Spartan-3E database took about 7\,000 ISE runs, each about a minute under emulation. Codeword batching
was decisive: measuring routing switches one per design would have needed hundreds of thousands of runs.
Two properties made it tractable: features combine by OR with few exceptions, and tile types repeat, so that
bits measured on one tile hold for every tile of the type and, across the family, for the other devices.

\paragraph*{Refusal of unmeasured features}
Because the router sees only measured switches and the writer refuses unmeasured features, a gap in the
database shows up as an error before a bitstream of unknown effect can reach a board.

\paragraph*{Own engine or nextpnr}
Writing a placer and router first gave a reference to measure nextpnr against, and the delay and packing
knowledge (pin delays, fanout loads, constants, inverters) that the architecture then needed. Moving to
nextpnr cost a few per cent of speed on the examples and register replication, and moved place and
route onto a widely used open engine that also supports other families.

\paragraph*{Running in a browser}
Keeping the flow in JavaScript and WebAssembly costs speed compared with native code, but removes
installation: the same flow runs on locked-down school computers, students' laptops and tablets.

\paragraph*{Limitations}
The shipped device descriptions are still derived from ISE's report. ISE's oracle checks bitstreams and
routing legality, not placement quality. The iCE40 flow has no such oracle yet. The Fast goal is slower than
ISE at its matching setting. The quality comparison covers one device (XC3S1200E) and one ISE version.

\section{Conclusion}
The flow produces vendor-identical bitstreams for two discontinued FPGA families and runs
entirely in a browser; with nextpnr placing and routing on architectures generated from observed data, it
builds smaller circuits than ISE at every goal and faster ones at Area and Speed. The database, the
experiments that measure it and the flow are open at \url{https://github.com/pmnmalo/Silinx}, and the
environment runs at \url{https://pmnmalo.github.io/Silinx/}. The toolchain is also a package of its own,
\texttt{silinx-fpga} 0.3.0 on npm (Apache-2.0, code and data), with a public API for Node.js and the browser,
taking a flat Verilog netlist or Yosys JSON as input.

\section*{Acknowledgment}
Silinx is not affiliated with or endorsed by AMD/Xilinx or Digilent; Xilinx, ISE, Spartan and Vivado
are trademarks of AMD. The code was generated with Claude Code (Anthropic) under the author's direction
and review.

\end{document}